# Observation of magnetization reversal in epitaxial $Gd_{0.67}Ca_{0.33}MnO_3$ thin films

Yanwei Ma[*],

*Applied Superconductivity Lab., Institute of Electrical Engineering, Chinese Academy of Sciences, Beijing 100080, China and CSIM, UMR 6511, CNRS- Universite de Rennes 1, Institut de Chimie, 35042 Rennes, France*

M. Guilloux-Viry, P. Barahona and O. Peña,

*CSIM, UMR 6511, CNRS- Universite de Rennes 1, Institut de Chimie, 35042 Rennes, France*

C. Moure

*Electroceramics Department, Instituto de Ceramica y Vidrio, CSIC, 28049 Madrid, Spain*

## Abstract

High–quality epitaxial thin films (~200 nm thick) of $Gd_{0.67}Ca_{0.33}MnO_3$ (GCMO) have been deposited onto (100) $SrTiO_3$ substrates by pulsed-laser deposition. Enhanced properties in comparison with bulk samples were observed. The magnetic transition temperature ($T_c$) of the as-grown films is much higher than the corresponding bulk values. Most interestingly, magnetization measurements performed under small applied fields, exhibit magnetization reversals below Tc, no matter whether the film is field-cooled (FC) or zero-field-cooled (ZFC). A rapid magnetization reversal occurs at 7 K when field cooled, while as for the ZFC process the magnetization decreases gradually with increasing temperatures, taking negative values above 7 K and changing to positive values again, above 83 K. In higher magnetic fields the magnetization does not change sign. The reversal mechanism is discussed in terms of a negative exchange *f-d* interaction and magnetic anisotropy, this later enhanced by strain effects induced by the lattice mismatch between the film and the substrate.

---

[*] Electronic mail: ywma@mail.iee.ac.cn

The mixed valent $A_{1-x}B_xMnO_3$ perovskite manganites, where A and B are rare-earth and divalent alkaline elements, have received a great amount of attention due to their unusual magnetic properties and colossal magnetoresistance effect (CMR) [1,2]. The parent compound $AMnO_3$ is well known to be an antiferromagnetic insulator, but becomes ferromagnetic metal upon doping. The theory of double exchange has been developed in order to explain this phenomenon and correctly predicts $x=1/3$ to be the optimal doping [3]. In addition, a second mechanism such as the strong electron-phonon interaction due to the Jahn-Teller effect is also required to explain the magnetoresistance within the double exchange model [4].

A large number of reports have described the structural, electrical and magnetic properties of pure and doped $AMnO_3$ perovksites; most of these reports concern, however, light rare-earths, that is, atoms with large ionic radii and of weak or non-existing magnetic nature (e.g., La, Pr, Nd…). Relatively, the perovskites based on heavy rare earths have not been studied well since their magnetic ordering temperatures are much smaller than those observed in, for instance, lanthanum-based perovskites. However, when incorporating heavy rare-earth elements, which have the largest magnetic moments of all the series, important modifications related to the magnetic response of the Mn sublattice and/or the overall magnetic behavior of the solid solution may be expected. Recently, reports on gadolinium-based perovskites have shown specific magnetic features, in particular a reversal of the magnetization at low temperatures when cooled in a magnetic field [5,6]. Similar anomalies were later found in praseodymium, neodymium and cerium compounds[7-10], in which we may find, for instance, an induced antiferromagnetic coupling of the Mn spins with the Ce spins, leading to a ferrimagnetic ground state [9].

On the other hand, the capability to fabricate manganites with unique physical properties into thin films makes it possible to create a rich variety of electronic and magnetic devices. It is also well known that films may present interesting physical properties, quite different from those of the materials produced by bulk ceramic techniques or single crystals with the same nominal composition, since lattice match between film and substrate is considered to be the most dominant factor for the epitaxial growth [11-13]. In this respect, we have undertaken the elaboration of thin films of $Gd_{0.67}Ca_{0.33}MnO_3$ (GCMO). The composition $x$=0.33 was chosen because this doping concentration should maximize the double exchange effect in a similar way as seen in the lanthanum compounds. More interestingly, we found multiple magnetization reversals below the ordering temperature in small applied fields, no matter whether the film is field-cooled or zero-field-cooled: a rapid magnetization reversal occurs at 7K when field cooled, while the magnetization reverses twice, at 7 and 83 K, during the ZFC process, a phenomenon which has not been reported yet. In this letter we report an investigation of the magnetic properties of GCMO thin films epitaxially grown on $SrTiO_3$ (100) substrates.

Thin films of $Gd_{0.67}Ca_{0.33}MnO_3$ (GCMO) were prepared by pulsed laser deposition (PLD) technique. The target, of nominal composition $Gd_{0.67}Ca_{0.33}MnO_3$, was a sintered pellet, of approximately 92 % of the theoretical density, prepared by a conventional solid-state reaction process. Films were synthesized on single-crystal (100) $SrTiO_3$ (STO) substrates. A detailed description of the deposition system is mentioned elsewhere [14]. In brief, a 248 nm KrF pulsed laser with 2 Hz repetition rate and 2 J/cm$^2$ energy density was used. Deposition was performed at 740°C under an oxygen pressure ranging between 0.25 and 0.6 mbar. Following the deposition, the films were cooled down to room temperature at a rate of about 35 °C/min in 200 Torr of oxygen. All films had a thickness around 200 nm.

The quality of the films produced was monitored by *in situ* reflection high energy electron diffraction (RHEED). The structural study was carried out by x-ray diffraction (XRD) using a high-resolution four-circle texture diffractometer (Brüker AXS D8 Discover) and CuK$_{\alpha 1}$ radiation ($\lambda$ = 1.54056 Å). The film microstructure was observed with a field effect scanning electron microscopy (SEM) (Jeol, JSM 6301F). Magnetization (M) was measured at various applied fields in a Quantum Design MPMS-XL5 SQUID magnetometer. Measured data of thin films contained contributions from the film and the STO substrate, this later being of negligible contribution. The applied fields were in the film plane. For each M(T) curve, the sample was cooled down to 4.2 K in zero magnetic field, and the magnetization was measured by warming the sample under an external field (ZFC). Then, the magnetization was obtained while the sample was being cooled under the same field (FC). The magnetization measurement was performed for several film samples to check reproducibility.

The streaky RHEED pattern that was evident throughout the GCMO film growth indicated that the growing film was at all times crystallized and smooth. GCMO films grown on STO substrates were found by XRD to be single phase with (00$l$) orientation, without any extra peaks due to impurities (Fig.1). In order to investigate the crystal quality of these films, the rocking curves of the (004) peaks were explored by $\omega$-scan. Typical full-width at half-maximum (FWHM) of all films obtained was comprised between 0.12° and 0.25°. Furthermore, the four peaks at 90° intervals in the $\varphi$-scan make evident the existence of an in-plane order of the films. These observations confirmed the high crystalline quality and good epitaxy of the GCMO thin films. In addition, the estimated value of the out-of-plane lattice parameter for the film was c/2=3.772 Å, almost identical to the lattice constant of the perovskite $Gd_{0.67}Ca_{0.33}MnO_3$ bulk material which was used as target in the PLD experiment ($a/\sqrt{2}$= 3.7914 Å, $b/\sqrt{2}$= 3.9447 Å, $c/2$=3.771 Å). As expected for epitaxial films, the in-plane lattice parameters $a$ and $b$ match the substrate of $SrTiO_3$ (cubic, $a$=3.905 Å) provided that the film crystal lattice is rotated by 45° with respect to the substrate. Therefore, the GCMO/STO films have the film-substrate lattice mismatch ~ 2.9% along $a$ direction, ~-1.0% along $b$ direction, with their in-plane lattice parameters expanded and compressed, respectively. Also, from the SEM observation, the GCMO film is uniform and dense. The film surface is covered with spherical grains with an average lateral size of 30 nm.

The typical zero-field cooled (ZFC) and field cooled (FC) magnetizations measured in a field of 20 Oe, as a function of temperature, are shown in Fig. 2. It is immediately seen that the ordering temperature Tc from a paramagnetic to a ferrimagnetic state is approximately 105 K, which is 25 K higher than that of the bulk of the same composition (Tc ~80 K) [5-6]. This difference in Tc is similar to what has been observed in $La_{1-x}Ba_xMnO_3$ thin films [13] and can be attributed to the lattice mismatch induced strains between the film and the substrate. From figure 2, it is observed that the data presents a large irreversibility between ZFC and FC magnetizations. During the FC process, the magnetization $M_{FC}$ increases rapidly below Tc, showing a large maximum at about $T_{cusp}^{FC}$ = 50 K. Upon further cooling, the magnetization decreases, intersects the temperature axis at a compensation temperature of $T_{comp}$= ~ 7 K and becomes negative. The temperature $T_{comp}$ is lower than that of the bulk $Gd_{0.67}Ca_{0.33}MnO_3$ ($T_{comp}$ ~15 K) [5-6]. This unusual phenomenon is named magnetization reversal, which indicates that below the compensation temperature, the magnetization inverts its sign and points opposite to the direction of the applied magnetic field [7-10].

During ZFC, that is, with increasing temperature from 5 K, the magnetization decreases gradually, crossing zero at approximately $T_{S1}$=~7 K and reaching a minimum negative value at T=~50 K. Upon further warming, $M_{ZFC}$ increases, crosses zero at a second compensation temperature of $T_{S2}$ ~ 83 K, reaching a small peak at $T_{cusp}^{ZFC}$ = 95 K, before decreasing gradually toward a paramagnetic state. It should be noted that the negative magnetization observed in the ZFC curve for our GCMO films (inset in Fig. 2) is much more important than whatever has been observed before in bulk samples of similar compositions. This difference may be ascribed to anisotropic effects, as it shall be discussed below. Such an anomaly is much more pronounced in low magnetic fields.

Figure 3 shows representative M-T curves measured under different applied fields between 20 and 5,000 Oe. Taking these curves into consideration, one can note the following features: (1) All FC curves tend to cross around $T_{comp}$=7 K over the whole range of the applied fields. However, for 5 kOe, the magnetization no longer becomes negative, but instead has a minimum at T = 7 K. (2) Applying a field above 250 Oe turns the ZFC magnetization to be positive in the whole temperature region (inset in Fig.3). (3) The two well-defined maxima observed in the magnetization exhibit quite different behaviors as a function of the applied field: while $T_{cusp}^{FC}$ stays at a rather constant value of 50 K, the other one (at $T_{cusp}^{ZFC}$) decreases with the applied field, going from 95 K (at H = 20 Oe) down to 50 K, the same value as $T_{cusp}^{FC}$. This situation is reversed to that observed in the GCMO bulk, where $T_{cusp}^{ZFC}$ stays constant at ~50 K, while $T_{cusp}^{FC}$ increases with increasing the field, until it reaches the same value as $T_{cusp}^{ZFC}$ [6]. In sufficiently large fields (5 kOe) there is no difference between the $M_{ZFC}$ and $M_{FC}$ curves.

In order to understand more deeply the origin of the magnetization reversal in the GCMO films, we have performed magnetization loops at temperatures comprised between 5 K and Tc, as illustrated in Fig. 4. These measurements were performed on a ZFC sample, under an external field varying between –20 kOe and +20 kOe. At a temperature of 5 K, the film is characterized by a coercive force Hc of 1.6 kOe ; the magnetization exhibits no saturation in fields of up to 50 kOe (not shown). The continuous increase of magnetization with the applied field indicates a ferrimagnetic-like state in the GCMO films.

As the temperature is increased to 20 K, the M-H behavior is quite different: the magnetization rises up dramatically at low field and tends to saturate at high field. The hysteresis loop has a coercivity of about 2.2 kOe, larger than that at 5 K. In contrast, the bulk of the same composition has a much smaller Hc at the same temperature [6]. With increasing temperature, the magnetization saturates more easily and the coercivity decreases quickly. At 90 K, temperature which almost corresponds to the peak of the ZFC curve, the M-H curve shows no hysteresis, but it is highly nonlinear, indicating a FM behavior. If the coercive field is plotted as a function of temperature (inset in Fig.4), there is a clear maximum in Hc(T) around 15 K. This behavior differs from that of the bulk of the same composition in which Hc decreases continuously from 5 K to Tc [6], and suggests that the unusual behavior of the ZFC magnetization may be attributed to associated domain effects or change in the magnetic anisotropy, as it will be discussed in more detail later.

Investigations of the electrical transport for this sample exhibit insulating like behavior (not shown). Below 150 K the resistance became too large to be measured using any of our methods. A similar result was also obtained by Snyder et al.: they reported that the GCMO samples show no transition to a metallic state down to 5 K as found in other manganites. Further, no extraordinary magnetoresistance was observed in the entire temperature range [5].

$Gd_{0.67}Ca_{0.33}MnO_3$ is considered to contain distinctive sublattices of gadolinium

and manganese, so that the magnetic state of the manganite samples is determined by the $f$–$f$, $f$–$d$, and $d$–$d$ exchange interactions involving $Gd^{3+}$, and both $Mn^{4+}$ and $Mn^{3+}$, all of which are magnetic. Magnetic interactions within the rare-earth sublattice are much weaker than the $d$–$d$ exchange between manganese ions. Furthermore, the replacement of gadolinium ions by nonmagnetic calcium ions must additionally decrease the $f$–$f$ exchange because of magnetic dilution. The Mn-Mn interactions in GCMO are considered to be of ferromagnetic character. Then, the magnetization reversal behavior should be due to an antiferromagnetic exchange coupling between the $Gd^{3+}$ and Mn sublattices, leading to a ferrimagnetic state due to a weaker, antiferromagnetic Mn-Gd interaction [5].

Under these assumptions, the above magnetic phenomena during FC can be easily understood as the existence of two magnetic sublattices, one aligned with the applied field and the other antialigned with the field. In a field-cooled measurement (FC) under small fields, as the temperature is decreased from Tc, the Mn sublattice orders ferromagnetically, creating a local field at the Gd site; at the same time, due to the negative $f$-$d$ exchange interaction, the Gd sublattice tends to align antiparallel to such local field. The driving force for this behavior is a molecular field ($H_M$), stemming from the Mn sublattice. The total magnetic moment ($M_{Mn}$-$M_{Gd}$) will reach a maximum at $T_{cusp}^{FC}$. If the antialigned sublattice magnetization grows more rapidly (e.g., proportional to $T^{-1}$) with decreasing temperature than the aligned one, the $T_{comp}$ is reached. Then the total magnetic moment becomes negative when $|M_{Gd}| > M_{Mn}$. The shape of the curve when field cooled is very similar to that for $Nd_{1-x}Ca_xMnO_3$ and $Dy_{0.67}Ca_{0.33}MnO_3$ [15-16], implying the same origin for both phenomena.

When the applied magnetic field is high enough, it overcomes the internal field produced by the Mn sublattice, and will predominate over the gadolinium susceptibility. The Gd ions remain parallel to the external field and the reversal of magnetization phenomenon is suppressed, although a strong dip in the magnetization is still observed at about 7 K. For instance, such is the case when the external magnetic field is 5 KOe or above (Fig.3).

On the other hand, the negative magnetization of the ZFC curve observed in the GCMO thin films is a quite unusual feature (Fig.2). Actually, multiple, reversible sign changes of the magnetization with temperature have been observed in many of the perovskite oxides [7, 17-20], but we are not aware of previous observations in the GCMO systems. A possible explanation for this behavior is the strong magnetic anisotropy associated with $Mn^{3+}$ ions occupying Jahn–Teller distorted $MnO_6$ octahedra, which may be caused by the structure distortion of our thin films, e.g. deformation and rotation of the $MnO_6$ octahedron due to large lattice mismatch between the GCMO film and the substrate STO. This scenario is strongly supported by a comparison of our data with magnetization measurements on bulk GCMO samples [6]. In the FC mode, the temperature dependence of the magnetization is qualitatively the same for the bulk and our thin film. However, in the ZFC mode, the bulk magnetization shows almost no negative values. This comparison strongly suggests that the magnetization reversal found in the ZFC process may result from the large magnetic anisotropy.

Indeed, during the ZFC process, the Mn magnetic domains are locked in random directions, giving rise to an AF canted state characteristic of these perovskite materials [21]. When an external field of 20 Oe is applied at low temperature (e.g. 5 K), the total magnetization ($M_{Mn}$-$M_{Gd}$) takes a small positive value because the Gd moments are not large enough to overcome the local field. When increasing the temperature from 5 K, the $M_{Gd}$ decreases more quickly than the $M_{Mn}$ magnetization. Since the measurement field is much smaller than the coercivity ($H_c$ = 1600 Oe at 5K), it is hard for the Mn domains to rotate. As a result, the total magnetization decreases, crossing zero (at approximately $T_{s1}$ = 7K), to a minimum negative value at T = 50 K. Due to the small coercivity above 50 K ($H_c$ = 100 Oe at 50K), the domains can gradually rotate to the external field direction ensuring that the total magnetization parallels the H direction. Meanwhile, the antialigned sublattice magnetization grows more slowly with increasing temperature than the aligned one. Hence, a quick increasing of magnetization from 50 K to about 95 K. At higher temperatures, the magnetic ordering is restrained, reaching the paramagnetic state and decreasing afterwards.

An alternative explanation may come from geometrical considerations, since the particles in a bulk material will tend to reorient and/or their spins will flip in the presence of an applied field in order to reach a stable state with moments aligned along the field direction. On the contrary, such phenomenon is much harder to occur in epitaxially grown thin films, since in addition to different grain sizes and grain-boundary conditions, the magnetic anisotropy at the boundaries is quite different from that of bulks because of lower symmetries and different crystal-field splitting.

In conclusion, we have studied the structure and magnetization of $Gd_{0.67}Ca_{0.33}MnO_3$ thin films grown onto (100) $SrTiO_3$ substrates by pulsed-laser deposition. The magnetic transition is remarkably sharp and the critical temperature of 105K is much higher than that of the bulk samples of the same composition. An original magnetic feature was found in the GCMO films under small applied fields, characterized by reversible changes of the signs of the magnetization with varying temperature, resulting in a temperature range where the magnetic moment is oriented in the direction opposite to the applied magnetic field. At sufficiently high magnetic fields the negative magnetic response is partially or fully suppressed. In a field-cooled measurement, if the antialigned sublattice magnetization grows more rapidly with decreasing temperature than the aligned one, an apparent diamagnetic state will result at sufficiently low temperatures. These results are interpreted as the occurrence of a negative exchange *f-d* interaction between the gadolinium sublattice and the $Mn^{3+}$-O-$Mn^{4+}$ ordered ferromagnetic sublatice. On the other hand, the negative magnetization observed in ZFC curves may be closely associated to geometrical effects related to the strong magnetic anisotropy caused by the structure distortion of the thin films which result from the large lattice mismatch between the GCMO film and the substrate STO. The results from these elementary substitutions may offer new insights for theories of the underlying physics in this class of compounds.

**Acknowledgments**

Financial support from Region Bretagne is greatly acknowledged.

**Captions**

Figure 1 $\theta$-$2\theta$ XRD patterns of $Gd_{0.67}Ca_{0.33}MnO_3$ films grown on STO substrates. The inset depicts the rocking curve of the (004) peak.

Figure 2 Temperature dependence of the magnetization for the GCMO thin film under ZFC and FC conditions, measured under 20 Oe. The inset shows an enlarged part of Fig.2.

Figure 3 Magnetization vs temperature in different magnetic fields, measured under ZFC (top panel) or FC (bottom panel) conditions. The inset shows an enlarged part of Fig.3 (top panel).

Figure 4 M-H loops measured between –2 T and +2 T at given temperatures, for a GCMO thin film. Inset: Temperature dependence of the coercive field (Hc), which has an anomalous peak near 15 K.

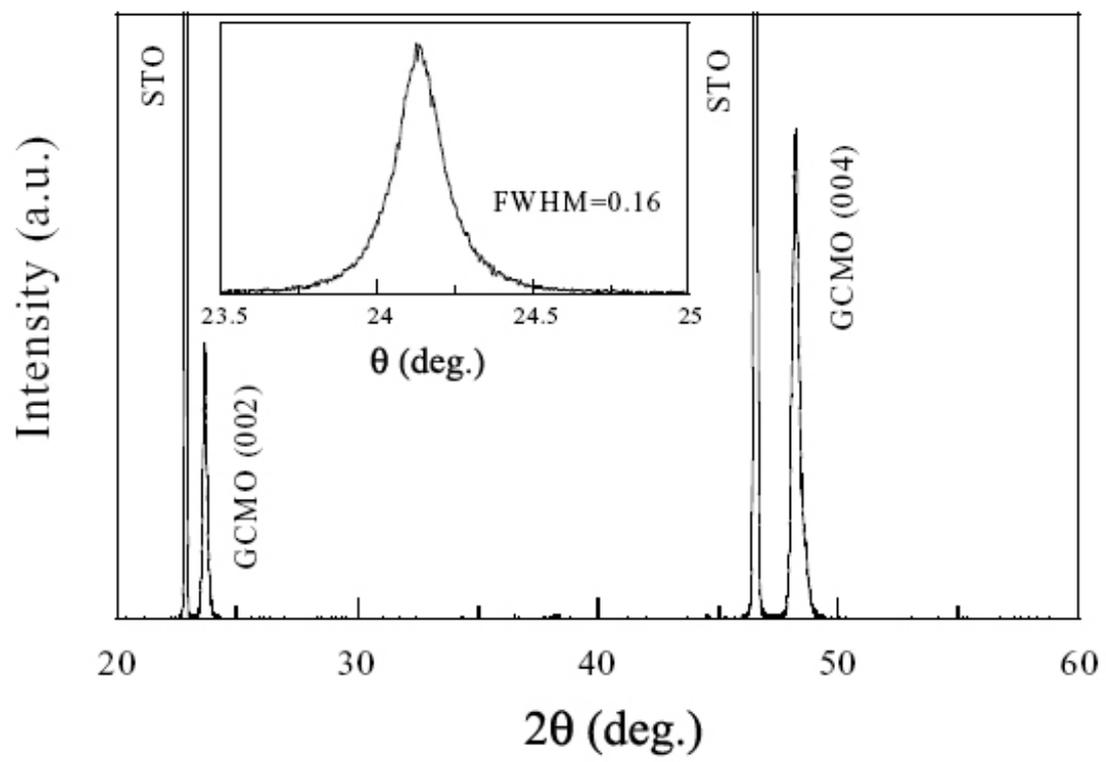

Fig. 1   Ma et al.

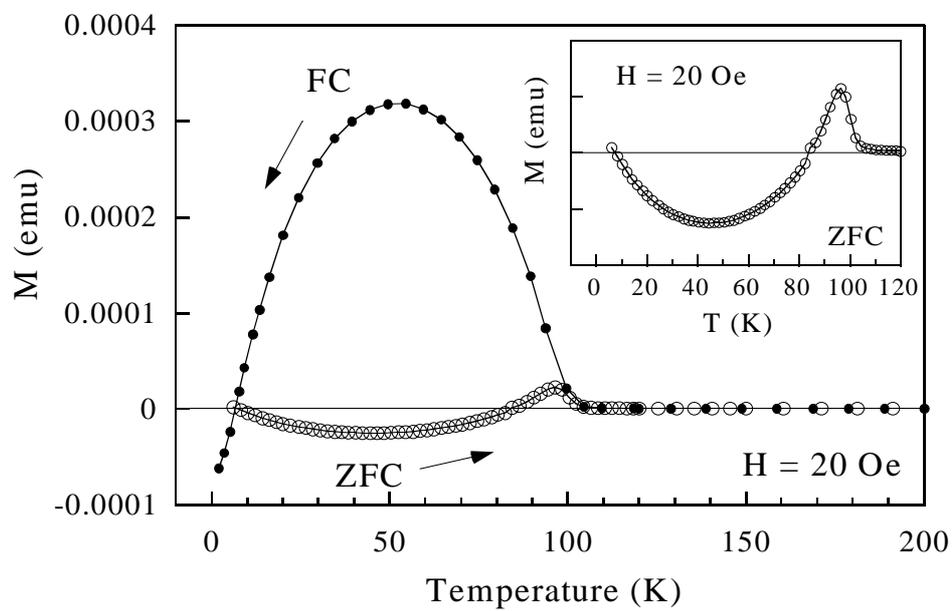

Fig.2  Ma et al.

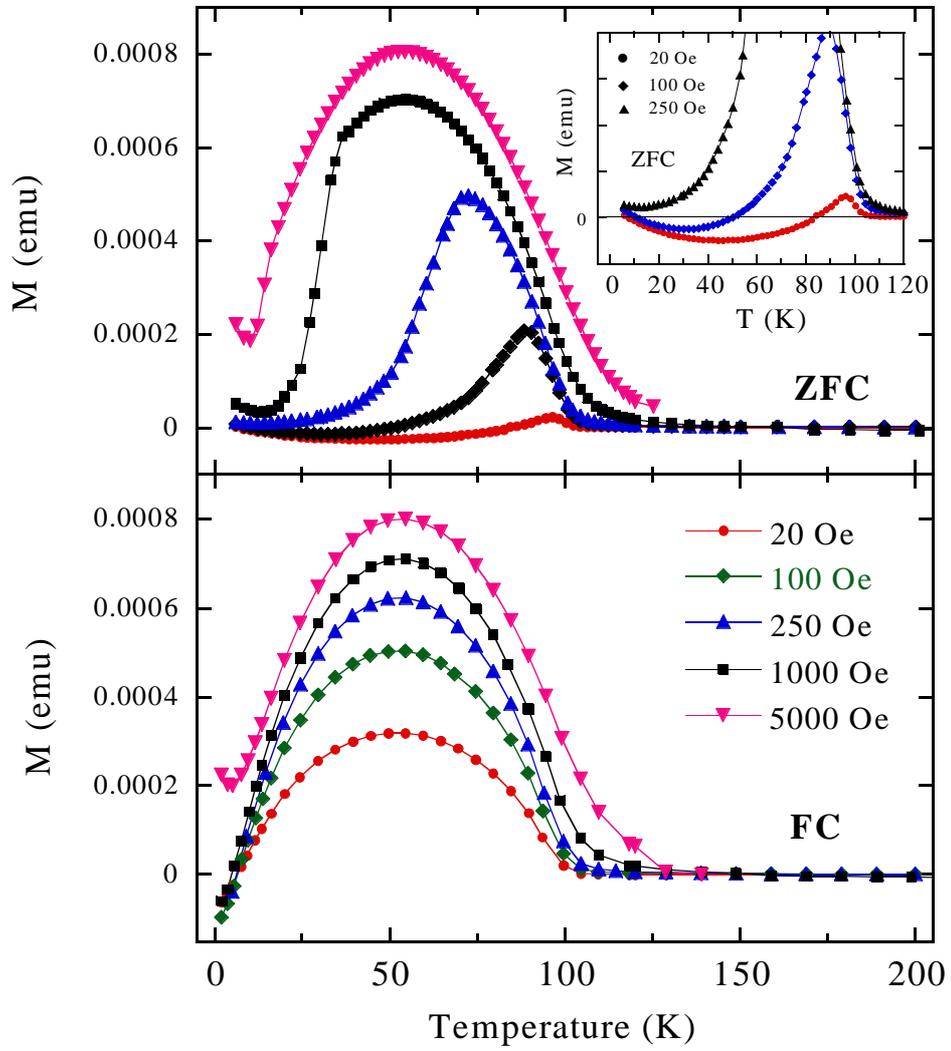

Fig. 3   Ma et al.

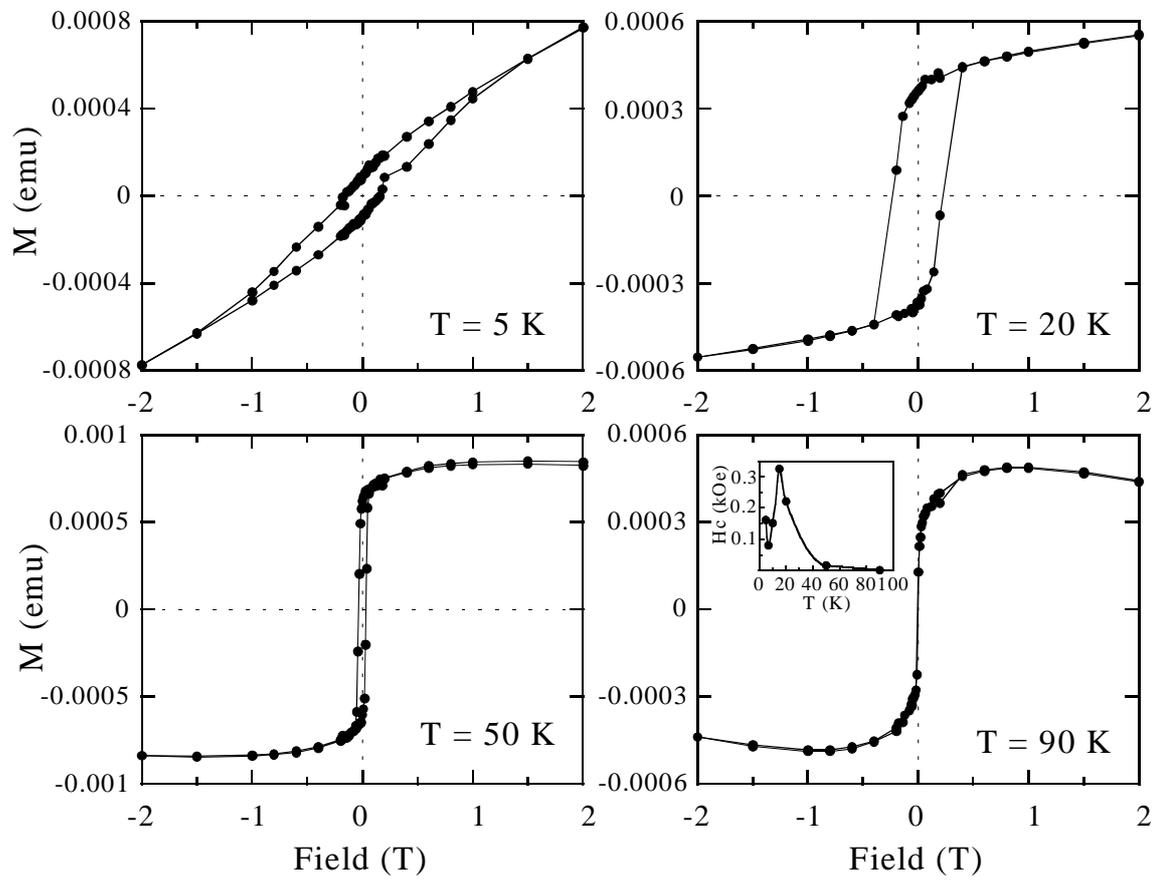

Fig. 4   Ma et al.